\documentclass[twocolumn,showpacs,preprintnumbers,amsmath,amssymb]{revtex4}
\usepackage{graphicx}
\usepackage{dcolumn}
\usepackage{bm}

\begin{document}

\preprint{APS/123-QED}

\title{Collapse of the N=28 shell closure in $^{42}$Si}
\author{B. Bastin$^2$, S. Gr\'evy$^{1*}$, D.
Sohler$^3$, O. Sorlin$^{1,4}$, Zs. Dombr\'adi$^3$, N.~L.
Achouri$^2$, J. C. Ang\'elique$^2$, F.~Azaiez$^4$,
D.~Baiborodin$^5$, R.~Borcea$^6$, C. Bourgeois$^4$, A. Buta$^6$, A.
B\"urger$^{7,8}$, R. Chapman$^9$, J.~C.~Dalouzy$^1$, Z.~Dlouhy$^5$,
A.~Drouard$^7$, Z. Elekes$^3$, S. Franchoo$^4$, S. Iacob$^6$, B.
Laurent$^2$, M.~Lazar$^6$, X.~Liang$^9$, E.~Li\'enard$^2$,
J.~Mrazek$^5$, L.~Nalpas$^7$, F. Negoita$^6$, N. A. Orr$^2$,
Y.~Penionzhkevich$^{10}$, Zs.~Podoly\'ak$^{11}$, F.~Pougheon$^4$,
P.~Roussel-Chomaz$^1$, M.~G.~Saint-Laurent$^1$, M.~Stanoiu$^{4,6}$
and I.~Stefan$^1$.} \affiliation{$^1$Grand Acc\'el\'erateur National
d'Ions Lourds (GANIL), CEA/DSM - CNRS/IN2P3, Bd Henri Becquerel, BP
55027, F-14076 Caen Cedex 5, France} \affiliation{$^2$Laboratoire de
Physique Corpusculaire, 6, bd du Mal Juin, F-14050 Caen Cedex,
France} \affiliation{$^3$Institute of Nuclear Research, H-4001
Debrecen, Pf.51, Hungary} \affiliation{$^4$Institut de Physique
Nucl\'eaire, IN2P3-CNRS, F-91406 Orsay Cedex, France}
\affiliation{$^5$Nuclear Physics Institute, AS CR, CZ-25068 Rez,
Czech Republic} \affiliation{$^6$Institute of Atomic Physics,
IFIN-HH, Bucharest-Magurele, P.O. Box MG6, Romania}
\affiliation{$^7$CEA Saclay, DAPNIA/SPhN, F-91191 Gif-sur-Yvette
Cedex, France} \affiliation{$^8$Helmholtz-Institut f\"ur Strahlen-
und Kernphysik, Universit\"at Bonn, Nu{\ss}allee 14-16, D-53115
Bonn, Germany} \affiliation{$^9$School of Engineering and Science,
University of Paisley, PA1 2BE Paisley, Scotland, UK}
\affiliation{$^{10}$FLNR, JINR, 141980 Dubna, Moscow region,
Russia}\affiliation{$^{11}$ University of Surrey, GU2 7XH Guildford,
United Kingdom }
\author{\vspace{-0.3cm}F. Nowacki$^{12}$ and A. Poves$^{13}$}
 \affiliation{$^{12}$IReS, BP28,
F-67037 Strasbourg Cedex, France} \affiliation{$^{13}$Departamento
de F\'isica Te\'orica, Universidad Aut\'onoma de Madrid, E-28049
Madrid, Spain.}

\date{\today}

\begin{abstract}

The energies of the excited states in very neutron-rich $^{42}$Si
and $^{41,43}$P have been measured using in-beam $\gamma$-ray
spectroscopy from the fragmentation of secondary beams of
$^{42,44}$S at 39 A.MeV. The low 2$^+$ energy of $^{42}$Si,
770(19) keV, together with the level schemes of $^{41,43}$P
provide evidence for the disappearance of the Z=14 and N=28
spherical shell closures, which is ascribed mainly to the action
of proton-neutron tensor forces. New shell model calculations
indicate that $^{42}$Si is best described as a well deformed
oblate rotor.
\end{abstract}

\pacs{23.20.Lv; 21.60.Cs; 27.40.+z; 29.30.Kv}

\maketitle

\indent Magic nuclei have in common a high energy for the first
excited state and a small transition probability B(E2: 0$^+_1
\rightarrow$ 2$^+_1$) compared to neighboring nuclei. This is
essentially due to the presence of large shell gaps, the origins and
configurations of which differ significantly along the chart of
nuclides. This implies a variable sensitivity of these shell gaps
with respect to the proton--neutron asymmetry. For instance, the
N=20 shell closure, bound by orbitals of opposite parity, $d_{3/2}$
below and $f_{7/2}$ above, remains remarkably rigid against
quadrupole deformation from $^{40}_{20}$Ca to $^{34}_{14}$Si even
after the removal of six protons~\cite{ibb98}. This feature can be
traced back to the large N=20 gap, the hindrance of 2$^+$
excitations due to the change of parity across it, and to the
presence of proton sub-shell gaps in the $sd$ shells at Z=16
($\sim$2.5~MeV) and Z=14 ($\sim$4.3~MeV)~\cite{dol76,tho84}.
Conversely, the N=28 shell closure, produced by the spin-orbit (SO)
interaction and separating the orbitals of same parity $f_{7/2}$ and
$p_{3/2}$, is progressively eroded below the doubly magic
$^{48}_{20}$Ca nucleus in $^{46}_{18}$Ar~\cite{gau06} and
$^{44}_{16}$S~\cite{gla97,gre05m}, after the removal of only two and
four protons, respectively. This rapid disappearance of rigidity of
the N=28 isotones has been ascribed to the reduction of the neutron
shell gap N=28 combined with that of the proton subshell gap Z=16,
leading to increased probability of quadrupole excitations within
the $fp$ and $sd$ shells for neutrons and protons, respectively. For
the $^{44}$S nucleus, its small 2$^+$ energy, large B(E2)
value~\cite{gla97} and the presence of a 0$^+_2$ isomer at low
excitation energy~\cite{gre05m} point to a mixed ground state
configuration of spherical and deformed shapes. As the proton Z=14
and neutron N=28 (sub)shell gaps have been proven to be effective in
$^{34}_{14}$Si and $^{48}$Ca$_{28}$ respectively, the search for a
new doubly magic nucleus would be naturally oriented towards
$^{42}_{14}$Si$_{28}$, which could be the lightest one with proton
and neutron gaps created by the SO interaction.\\
\indent The experimental status concerning the structure of
$^{42}$Si is rather controversial. On the one hand the very short
$\beta$-decay lifetimes of the $^{40-42}_{\;\;\;\;\;\;14}$Si nuclei
point to deformed ground state configurations in the Si isotopic
chain~\cite{gre04}. On the other hand the weak two proton knockout
cross section $\sigma_{-2p}(^{44}$S$ \rightarrow^{42}$Si) was given
as a strong indication in favor of a "doubly magic" spherical
nucleus~\cite{fri05} with a large Z=14 shell gap. However, the same
authors have shown more recently~\cite{fri06} that a reduction of
the Z=14 gap by as much as 1~MeV does not increase the
$\sigma_{-2p}$ value significantly. Further, the newly measured
atomic mass of $^{42}$Si~\cite{jur06} is compatible with an excess
of microscopic energy as compared to a spherical liquid drop, which
could be obtained either for a spherical or deformed shell closure.\\
\indent The present work aimed at determining the 2$^+$ energy of
$^{42}_{14}$Si$_{28}$ and the energy spectrum of the neutron-rich
$^{41,43}_{\;\;\;\;15}$P$_{26,28}$ isotopes to obtain a global
understanding of the proton and neutron excitations at Z=14 and N=28
to infer whether $^{42}$Si can be considered as a new doubly magic
nucleus or not.

\begin{figure}
\includegraphics[width=8.5cm]{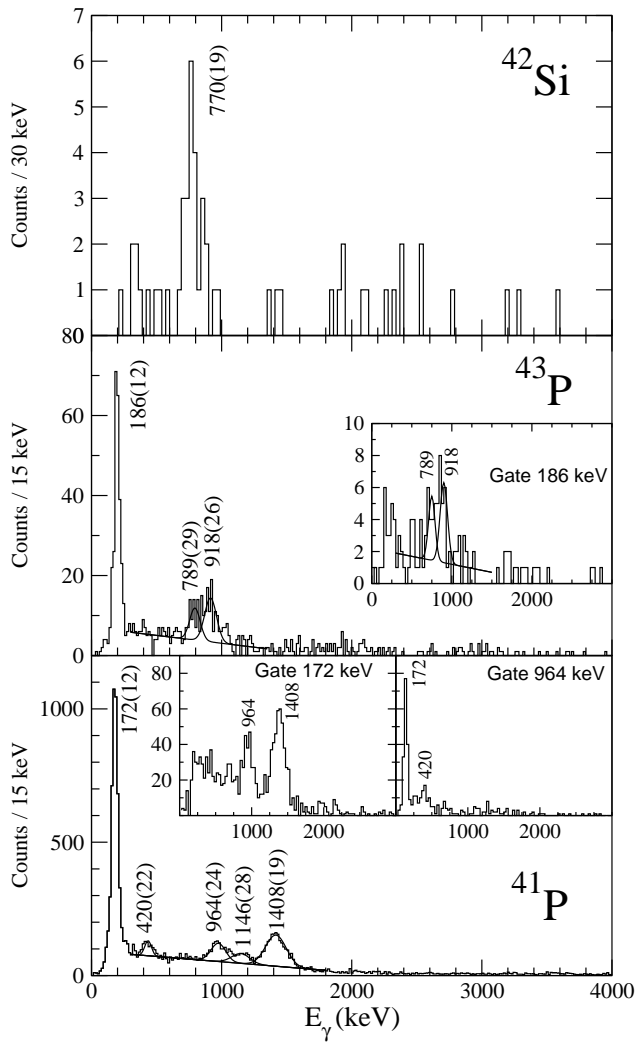} \caption{\label{fig:gamma}
$\gamma$-ray spectra observed in coincidence with the $^{42}$Si
(upper), $^{43}$P (middle) and $^{41}$P (bottom) nuclei. In the
insets $\gamma$-$\gamma$ coincidence spectra are presented, gated
with transitions belonging to the $^{41,43}$P nuclei.}
\end{figure}

\begin{figure}
\includegraphics[width=8.5cm]{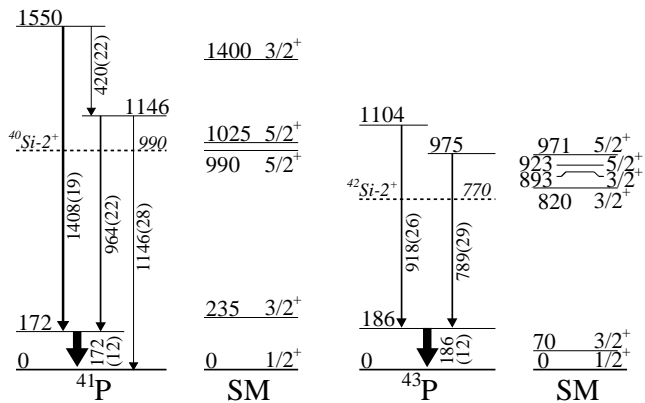}
\caption{\label{fig:41p_43p_ls} Level schemes for $^{41,43}$P and
the corresponding shell model calculations (SM) performed in the
present work. Positions of the 2$^+$ states in $^{40,42}$Si are
shown by dashed lines.}
\end{figure}

\indent The experiment was carried out at the Grand Acc\'el\'erateur
National d'Ions Lourds ({\sc GANIL}) facility. A primary beam of
$^{48}$Ca at 60~A$\cdot$MeV impinged onto a 200~mg/cm$^2$ C target
with a mean intensity of 3.8~$\mu$A to produce a cocktail of
projectile like fragments. They were separated by the {\sc ALPHA}
spectrometer, the magnetic rigidity of which was set to optimize the
transmission of the $^{44}$S nuclei, produced at a rate of
125~s$^{-1}$. Fragments were guided over a flight path of about 80~m
along which the time-of-flight (TOF) was determined with two
microchannel plates. A thin 50~$\mu$m Si detector was used to
determine the energy losses ($\Delta$E), which completed the event
by event identification prior to the fragmentation in a secondary
target of 195~mg/cm$^2$ of $^9$Be. This target was placed at the
entrance of the SPEG spectrometer,
the magnetic rigidity of which was tuned to maximize the
transmission of the $^{42}$Si nuclei produced through a 2 proton
knockout reaction. At the focal plane of SPEG, ionisation and
drift chambers provided information on the $\Delta$E and positions
of the transmitted nuclei, while a plastic scintillator was used
to determine TOF and residual energies. The $\sigma_{-2p}(^{44}$S
$\rightarrow ^{42}$Si) cross section is measured to be
80(10)~$\mu$barn at 39~A$\cdot$MeV. This agrees with the value of
120(20)~$\mu$barn obtained at 98.6~A$\cdot$MeV~\cite{fri05}, after
having taken into account a enhancement factor of 25\% due to the higher
beam energy\cite{tos07}. \\
\indent An array of 74~BaF$_2$ crystals, each of 9~cm diameter and
14~cm length, was arranged in two hemispheres above and below the
Be target at a mean distance of 25~cm to detect $\gamma$-rays
arising from the secondary reactions. The energy threshold of the
detectors was around 100~keV. The energies of two $\gamma$-rays
detected in coincidence in adjacent detectors were combined by
add-back.  The $\gamma$-ray energies were corrected for Doppler
shifts due to the in-flight emission by the fragments. Photopeak
efficiencies of 38~\% at 779~keV, 24~\% at 1.33~MeV and 16~\% at
2~MeV were achieved for fragments with $v/c\simeq 0.3$. The energy
resolution was 15\% at 800~keV and 12\% at 1.4~MeV, which includes
the effects of the intrinsic resolution of the detectors and the
Doppler broadening. Using the systematics of the peak widths
deduced from the observation of known single peaks, doublets of
$\gamma$-rays could also be disentangled. The time resolution of
the array was 800~ps. This enabled a clean separation between
neutrons or charged particles and the $\gamma$-rays arising from
the reaction on the basis of their TOF differences.\\
\indent $\gamma$-ray spectra obtained for the $^{42}$Si and
$^{41,43}$P isotopes are shown in Fig.~\ref{fig:gamma}. Several
other nuclei were also produced, such as $^{38,40}$Si and $^{44}$S.
Noteworthy is the fact that we find good agreement with their
previously measured $\gamma$ transitions, published in
Refs.~\cite{cam06,gla97}. A clear single peak is visible in the
spectrum of $^{42}$Si at 770(19)~keV on a significance level of more
than 3 $\sigma$. It can be assigned to the decay of the first
excited state, namely the 2$^+_1\rightarrow$0$^+_1$ transition. The
number of counts corresponds to a feeding of the 2$^+$ state of
44$\pm$10\%. In $^{43}$P, in addition to the low energy transition
at 186~keV previously reported by Fridmann~{\it et
al.}~\cite{fri05}, a doublet of weak transitions is visible at 789
and 918~keV. Both transitions are in coincidence with the 186~keV
transition, and are placed on top of it in the proposed level scheme
(Fig.~\ref{fig:41p_43p_ls}). In $^{41}$P, similarly to the $^{43}$P
case, a strong low energy transition at 172(12)~keV is visible
together with several slightly overlapping peaks at 420, 964, 1146
and 1408~keV. The 964 and 1408~keV transitions are in coincidence
with the 172~keV transition, while the 964~keV transition is in
coincidence with that at 420~keV too. In addition to coincidence
relationships, intensity arguments are used in order to build the
level scheme presented in Fig.~\ref{fig:41p_43p_ls}. The present
observation of {\it new} $\gamma$-lines in the $^{43}$P and
$^{42}$Si is explained by an enhanced $\gamma$ efficiency by about a
factor of ten at 1 MeV with respect to the work of
Refs.\cite{fri05,fri06}.

\begin{figure}
\includegraphics[width=8cm]{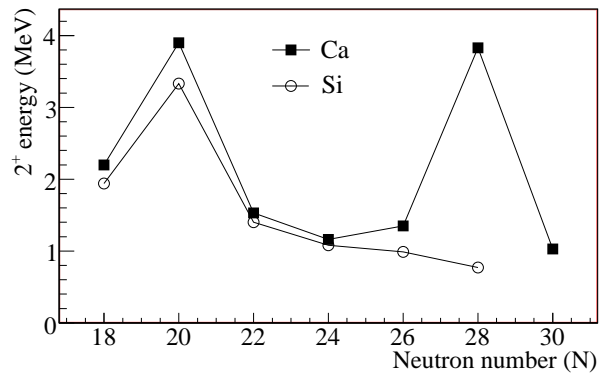}
\caption{\label{fig:energies} Energies of the 2$^+$ states measured
in the Ca and Si isotopes. Present result for $^{42}$Si --770(19)
keV-- brings evidence for the the collapse of the N=28 shell closure
at Z=14.}
\end{figure}

Fig.~\ref{fig:energies} shows the energies of the 2$^+$ states in
the Si and Ca isotopes. The correlated increase of the 2$^+$
energies in the $^{40}$Ca and $^{34}$Si nuclei at N=20 does not hold
at N=28. While 2$^+$ energies increase in Ca isotopes from N=24 to
reach a maximum around 4 MeV at N=28, the 2$^+$ energies in the Si
isotopes start to deviate from those of the Ca at N=26, reaching a
minimum value at N=28. The slight deviation from the Ca curve at
N=26 was interpreted in Ref.~\cite{cam06} as an indication in favor
of a reduced N=28 shell gap. However, neutron excitations occurring
{\it before} the complete filling of the $\nu f_{7/2}$ shell are not
especially sensitive to the N=28 gap, as the 2$^+$ states are
qualitatively due to excitations {\it inside} the $f_{7/2}$ shell.
Conversely, in $^{42}$Si$_{28}$ the 2$^+$ state comes from
particle-hole excitations across the gap and therefore, the dramatic
decrease of its 2$^+$ energy leaves no doubt concerning the
disappearance of the spherical N=28 shell closure at Z=14, the
energy of 770~keV being one of the smallest among nuclei having a
similar mass. It is worth pointing out, as shown in
Fig.~\ref{fig:41p_43p_ls}, that the decrease of the $2^+$ energies
in the $^{40,42}$Si nuclei is correlated to the behavior of those
states observed around 1 MeV in the $^{41,43}$P isotones, which in
the shell model arise from the coupling of the last proton in the
$s_{1/2}$ or $d_{3/2}$ orbitals to the $2^+$ excitation.
This provides additional support to the disappearance of the N=28
spherical gap. If the N=28 gap had persisted in the P isotopes, a
sequence of levels similar to those in $^{47}$K would have been
observed. Namely two nearby 1/2$^+$ and 3/2$^+$ states originating
from the quasi-degeneracy of the $\pi s_{1/2}$ and $\pi d_{3/2}$
orbitals and a large gap in energy ($\sim$2MeV) above this doublet.

The spectroscopy of the nuclei with proton number in the range of 16
to 20 and neutron number from 20 to 28 is reproduced successfully
when using large scale shell model calculations in a valence space
comprising the full $sd$-shells for protons and $pf$-shells for
neutrons with the effective interaction SDPF-NR~\cite{cau04}. The
remarkable features of this interaction which account for the
evolution of the nuclear structure between N=20 and N=28 are : (i)
the decrease of the $d_{3/2}$-$s_{1/2}$ proton splitting by about
2.5~MeV from $^{39}$K$_{20}$ to $^{47}$K$_{28}$ as the neutron
$f_{7/2}$ orbital is filled, (ii) the reduction of the N=28 gap by
about 330~keV per pair of protons~\cite{gau06} removed from the $^{48}$Ca.\\
\indent Despite these successes, a correct description of the Si
isotopes cannot be straightforwardly obtained by using the monopole
matrix elements of the SDPF-NR interaction derived for the Ca
isotopes. This arises from the fact that the occupied and valence
proton orbitals involved in both cases are somewhat different. In
particular, some amount of core excitations to the $\pi f_{7/2}$
orbital, which lies just above the valence space of the Ca isotopes,
is included in the neutron-pairing matrix elements $V^{nn}$ of
$pf$-shell orbits. This contribution is expected to be negligible in
the Si isotopes, in which the $\pi f_{7/2}$ orbital lies at much
higher excitation energy. According to the model, the configuration
of the 2$^{+}$ states in $^{36,38,40}$Si arises mainly from
pair-breaking inside the $\nu f_{7/2}$ shell and therefore the 2$^+$
energies scale with the $V^{nn}$ values. Indeed the energy of the
2$^{+}$ states in $^{36,38,40}$Si~\cite{ibb98,lia06,cam06} and those
of the $5/2^+$ states in $^{37,39}$P~\cite{sor04} are overestimated
by the SDPF-NR interaction up to 300-400~keV. The discrepancy is
larger for $^{42}$Si, the calculated 2$^{+}$ energy of which being
1.49~MeV, instead of 770~keV. Therefore a reduction of the $pf$
shell $V^{nn}$ matrix elements by 300~keV gives the 2$^{+}$ energies
of $^{36,38,40}$Si as well as the $5/2^+$ energies in
$^{37,39}$P~\cite{sor04} in agreement with their experimental
values. Nevertheless the calculated 2$^{+}$ energy in $^{42}$Si,
1.1~MeV, is still larger than the experimental value. To further
reduce it without changing the properties of the other N=28
isotones, the proton-neutron monopole matrix elements
$V_{d_{5/2}(fp)}^{pn}$ can be considered. Indeed, the $d_{5/2}$
orbital is "active" in the Si isotopes whereas it is too deeply
bound to play a significant role in the description of the Ca
isotopes. Nevertheless, with the SDPF-NR interaction the proton
$d_{3/2}$-$d_{5/2}$ splitting in $^{39}$K and $^{47}$K overestimates
the experimental values, see Tab.~\ref{tab:1}. By modifying
adequately the $V_{d_{5/2}(fp)}^{pn}$ matrix elements, the
$d_{3/2}$-$d_{5/2}$ splitting can be better adjusted in the K
isotopes (Tab.~\ref{tab:1}) and therefore the 2$^+$ energies in
$^{36,38,40,42}$Si, as well as the level schemes of $^{41,43}$P,
agree nicely with the experimental data shown in
Figs.~\ref{fig:41p_43p_ls} and \ref{fig:energies}. This modification
leads to a decrease of the $d_{3/2}$-$d_{5/2}$ splitting by 1.94~MeV
from $^{34}$Si to $^{42}$Si which results in an enhanced
collectivity. $^{42}$Si becomes clearly an oblate rotor up to
$J$=8$^+$; with a calculated 2$^{+}$ excitation energy of 810~keV
and an intrinsic quadrupole moment $Q_i$ of --87~e~fm$^2$
corresponding to a quadrupole deformation $\beta$=-0.45. The doubly
magic (N=28, Z=14) component is only 20\% of the ground state
wavefunction with an average 2.2 neutrons above N=28 and 1.2 protons
above Z=14; the percentage of the closed proton configuration
(0d$_{5/2})^6$ being 33\%. With a present Z=14 gap of 5.8~MeV, our
measured $\sigma_{-2p}(^{44}$S$ \rightarrow^{42}$Si) cross section
agrees with the one calculated with a similar gap in
Ref.~\cite{fri06}.

\begin{table}
\caption{Proton $d_{3/2}$-$d_{5/2}$ splitting in $^{39}$K and
$^{47}$K} \label{tab:1}
\begin{ruledtabular}
\begin{tabular}{lll}\noalign{\smallskip}
& $^{39}$K &$^{47}$K \\
\noalign{\smallskip}\hline\noalign{\smallskip}
experiment& 6.74~\cite{dol76} & 4.84~\cite{kra01}\\
shell model~\cite{cau04} & 7.4 & 5.92\\
shell model (this work)  & 7.18 & 4.93\\
\noalign{\smallskip}
\end{tabular}
\end{ruledtabular}
\end{table}
\indent The resulting shell model description of the N=20 to N=28
region south of Ca isotopes is the following: as compared to the
$^{34}$Si and $^{48}$Ca nuclei, major changes in the energies of the
proton and neutron orbitals are occurring towards $^{42}$Si. The
complete filling of the neutron $f_{7/2}$ shell from $^{34}$Si leads
to a shrinkage of the $sd$ orbitals, with a \emph{proton} SO
splitting $d_{3/2}$-$d_{5/2}$ reduced by about 1.94~MeV. This
reduction is compatible with what is accounted for by tensor forces,
which act attractively (repulsively) between the proton and neutron
orbits $d_{3/2}$-$f_{7/2}$ ($d_{5/2}$-$f_{7/2}$) as described in
Ref.~\cite{ots05,gad06}. Simultaneously, as depicted in
Ref.~\cite{gau06} the removal of protons from
$^{48}$Ca induces a compression in energy of the four \emph{neutron}
$fp$ orbits and hence a reduction of the N=28 gap by about 1~MeV in
$^{42}$Si (from an initial value of 4.8 MeV in $^{48}$Ca) due to the
combined effects of the proton-neutron tensor force and the density
dependence of the SO interaction. The overall picture would be then
that the mutual actions of the proton-neutron tensor forces in
$^{42}$Si induce the reduction of the neutron N=28 gap and limit the
size of the proton Z=14 gap. In addition, particle-hole excitations
between occupied and valence orbitals which are separated by $\Delta
\ell , j$=2 for both protons ($\it sd$) and neutrons ($\it fp$)
naturally favor quadrupole correlations that generate collectivity
through mechanisms related to Elliott's SU(3) symmetry~\cite{ell56}.
The combined effects of the compression of the proton and neutron
orbitals, plus the quadrupole excitations, produce a rich variety of
behaviors and shapes in the even N=28 isotones; spherical $^{48}$Ca;
oblate non-collective $^{46}$Ar; coexistence in $^{44}$S, and two
rotors, oblate $^{42}$Si and prolate $^{40}$Mg. This variety of
shapes is also globally supported by mean field calculations,
relativistic or
non-relativistic~\cite{wer96,rei99,lal99,per00,rod02}.

%
\indent To summarize, the energies of excited states in the
$^{42}$Si and the $^{41,43}$P nuclei have been measured through
in-beam $\gamma$-ray spectroscopy. The low energy of the 2$_1^+$
state in $^{42}$Si, 770(19) keV, together with the level schemes of
$^{41,43}$P provide evidence for the disappearance of the N=28
shell-closure around $^{42}$Si. It is ascribed to the combined
action of proton-neutron tensor forces leading to a global
compression of the proton and neutron single particle orbits, added
to the quadrupole symmetry between the occupied and valence states
which favors excitations across the Z=14 and N=28 shell gaps.

\indent This work benefits from discussions with A.~Navin, J.
Tostevin and L. Gaudefroy. The authors are thankful to the GANIL and
LPC staffs. We thank support by BMBF 06BN109, GA of Czech Republic
202/040791, MEC-DGI-(Spain) BFM2003-1153, and by the EC through the
Eurons contract RII3-CT-3/2004-506065 and OTKA T38404-T42733-T46901.
RB, AB, ML, SI, FN and RC, XL acknowledge the IN2P3/CNRS and EPSRC
support.

\indent $^*$ corresponding author: grevy-at-in2p3.fr \\

\end{document}